\documentclass[nocompress]{spie}  

\usepackage{amsmath,amsfonts,amssymb}
\usepackage{graphicx}
\usepackage[colorlinks=true, allcolors=blue]{hyperref}

\title{Designing Metasurfaces to Manipulate Antenna Radiation}

\author[a]{James R. Capers*}
\author[b]{Stephen J. Boyes}
\author[a]{Alastair P. Hibbins}
\author[a]{Simon A. R. Horsley}
\affil[a]{Centre for Metamaterial Research and Innovation, University of Exeter, Stocker Road, Exeter, EX4 4QL, UK}
\affil[b]{DSTL Porton Down, Salisbury, Wiltshire, SP4 0JQ, UK}

\authorinfo{*Email: jrc232@exeter.ac.uk}

\begin{document} 

\maketitle

\begin{abstract}
    Designer manipulation of light at the nanoscale is key to several next--generation technologies, from sensing to optical computing.
    One way to manipulate light is to design a material structured at the sub--wavelength scale, a metamaterial, to have some desired scattering effect.
    Metamaterials typically have a very large number of geometric parameters than can be tuned, making the design process difficult.
    Existing design paradigms either neglect degrees of freedom or rely on numerically expensive full--wave simulations.
    In this work, we derive a simple semi--analytic method for designing metamaterials built from sub--wavelength elements with electric and magnetic dipole resonances.
    This is relevant to several experimentally accessible regimes.
    To demonstrate the versatility of our method, we apply it to three problems: the manipulation of the coupling between nearby emitters, focusing a plane wave to a single point and designing a dielectric antenna with a particular radiation pattern.
\end{abstract}


\section{INTRODUCTION \label{sec:intro}}

The desire to manipulate visible light has existed for well over 2000 years \cite{Smith2015}.
Research on this topic has borne several technologies key to modern life, from spectacles to fibre optical cables.
For at least the last two decades, technology trends have pushed for ever greater miniaturisation, performance and efficiency.
One proposed solution to these challenges is to build computing elements from optical devices.
Recent demonstrations of this principle include optical differentiation for edge detection \cite{Zhou2020}, optical integration \cite{Ferrera2010}, systems that solve differential equations \cite{Tan2013} and optical neural networks \cite{Shastri2021}.
At a fundamental level all of these technologies require the ability to manipulate light at the nanoscale in designer ways.  
One way to achieve this is to take inspiration from radio frequency applications, where waves of different frequency and polarisation are regularly re--directed or re--shaped by antenna systems.  
In the same way, optical light can be controlled by resonant metallic structures \cite{Hulst2017, Giannini2011}, `plasmonic antennas'.

Plasmonic antennas, typically build from a small number ($\sim 2-10$) of resonant metallic elements, suffer from a couple of significant drawbacks.
Firstly, plasmonic structures have large absorption at optical wavelengths, limiting efficiency.
An alternate approach is to use dielectric resonators rather than plasmonic ones \cite{Staude2017}.
Dielectric resonators exhibit shape dependant Mie resonances \cite{Mie1908} that have lower loss at optical wavelengths than plasmonic resonances.
Secondly, due to the small number of elements, the number of degrees of freedom is limited.
This can make it difficult to design plasmonic antennas that have arbitrary effects upon light.
To achieve more general control of light, one can assemble structures made from several ($\gg 10$) plasmonic or dielectric elements \cite{Meinzer2014}, giving many more design degrees of freedom.
Built from several discrete sub--wavelength elements, this kind of structure is a metamaterial.
As metamaterials have many geometric parameters that can be tuned to change the response of the material to electromagnetic waves, finding a set of parameters that give a particular response can be very challenging.

Many methods to solve this `inverse design' problem have emerged recently \cite{Molesky2018}.
To design metasurface lenses \cite{Khorasaninejad2016} or holograms \cite{Ni2013}, where the function of the metamaterial is to impart a known phase offset onto the incident field, the Gerchberg--Saxton algorithm \cite{GS1972} is commonly used.
This algorithm is simple and efficient, however assumes that elements of the metasurface do not strongly couple to each other and typically requires many full--wave simulations to build up a library of the many phase--changing elements from which the metasurface is built.
The design of aperiodic metamaterials built from discrete resonating elements, with the aim of coupling to emitters, can be facilitated with genetic algorithms \cite{Wiecha2018a, Wiecha2018b}. 
Genetic algorithms are extremely effective at exploring large and complex search--spaces with many local minima, however the resulting structures can be difficult to understand intuitively \cite{Yeung2020}. 
Topology optimisation \cite{TObook} is used extensively to design graded index structures for a wide range of functionality including wavelength splitters\cite{Piggott2015}, lenses that overcome the diffraction limit \cite{Otomori2017} and  mode sorters \cite{Frellsen2016}.
The optimisation is usually performed using gradient descent, which can be slow particularly for large structures or fine discretisations.
One way to accelerate this is to make use of reciprocity \cite{LL8} to convert the slow calculation of a gradient into two field calculations.
This `adjoint' method \cite{Miller2012, Keraly2013} can be very efficient, however still requires many full--wave simulations over the course of the iterative optimisation.

In this work, we derive a method for designing metamaterials made from discrete scatterers.
By assuming that the scatterers support only electric and magnetic dipole resonances, valid for small scatterers at optical wavelengths \cite{Kuznetsov2016}, Maxwell's equations can be solved exactly eliminating the need for full--wave simulations.
These solutions are developed in Section \ref{sec:dda}.
To achieve the required scattering properties, the desired figure of merit can be expanded under small changes in the position of a scatterer.
This gives an analytic expression that can be used to iteratively update the scatterer locations to maximise or minimise the figure of merit.
We derive and apply this procedure to several relevant problems in Section \ref{sec:designing}, including manipulating the coupling between two nearby emitters, focusing a plane wave to a point and designing a structure with a particular radiation pattern.

\section{THE DISCRETE DIPOLE APPROXIMATION}
\label{sec:dda}

To address the problem of designing metasurfaces, we begin by considering a metasurface composed of sub--wavelength discrete elements that support electric and magnetic dipole resonances.
Maxwell's equations for a fixed frequency $\omega = ck$, where $k$ is the wave--number, can then be written as 
\begin{equation}
    \begin{pmatrix}
        \nabla \times \nabla \times & 0 \\
        0 & \nabla \times \nabla \times
    \end{pmatrix}
    \begin{pmatrix}
        \boldsymbol{E} \\
        \boldsymbol{H}
    \end{pmatrix}
    - k^2
    \begin{pmatrix}
        \boldsymbol{E} \\
        \boldsymbol{H}
    \end{pmatrix}
    = 
    \begin{pmatrix}
        \boldsymbol{E}_s \\
        \boldsymbol{H}_s
    \end{pmatrix}
    + 
    \begin{pmatrix}
        \omega^2 \mu_0 & i \omega \mu_0 \nabla \times \\
        -i\omega \nabla \times & k^2
    \end{pmatrix}
    \begin{pmatrix}
        \boldsymbol{P} \\
        \boldsymbol{M}
    \end{pmatrix} .
    \label{eq:maxwell}
\end{equation}
In this expression $\boldsymbol{E}_s$ and $\boldsymbol{H}_s$ represent the source fields, for example the field due to an emitter or a background plane wave.
The properties of the metasurface are encoded in the polarisation density $\boldsymbol{P}$ and the magnetisation density $\boldsymbol{M}$.
This is generally a difficult equation to solve, however the assumption that the scatterers are sub--wavelength $r k \leq 1$ means that the elements of the metasurface can be modelled as point--like.
In general, the polarisation and magnetisation densities contain all multipole moments \cite{Raab2005, Evlyukhin2011, Evlyukhin2013}, however if we assume that only the dipole terms are present, then we can write the polarisation and magnetisation densities as 
\begin{align}
    \boldsymbol{P} &= \sum_n \boldsymbol{\alpha}_E \boldsymbol{E} (\boldsymbol{r}_n) \delta (\boldsymbol{r}-\boldsymbol{r}_n), &
    \boldsymbol{M} &= \sum_n \boldsymbol{\alpha}_H \boldsymbol{H} (\boldsymbol{r}_n) \delta (\boldsymbol{r}-\boldsymbol{r}_n) .
    \label{eq:PM}
\end{align}
This reduces the source terms in Maxwell's equations (\ref{eq:maxwell}) to a summation of delta functions.
Equations of this form can be solved with the dyadic Greens function and its curl \cite{Schwinger1950, Tai1993}
\begin{align}
    \boldsymbol{G} (\boldsymbol{r}, \boldsymbol{r'}) &= \left[ \boldsymbol{1} + \frac{1}{k^2} \nabla \otimes \nabla \right] \frac{e^{ik|\boldsymbol{r}-\boldsymbol{r'}|}}{4 \pi |\boldsymbol{r}-\boldsymbol{r'}|}, &
    \boldsymbol{G}_{EH} (\boldsymbol{r}, \boldsymbol{r'}) &= \nabla \times \boldsymbol{G} (\boldsymbol{r}, \boldsymbol{r'}) .
\end{align}
The solution to Maxwell's equations (\ref{eq:maxwell}) with source terms given by (\ref{eq:PM}) can then be written as
\begin{equation}
    \begin{pmatrix}
        \boldsymbol{E} (\boldsymbol{r}) \\
        \boldsymbol{H} (\boldsymbol{r})
    \end{pmatrix}
    = 
    \begin{pmatrix}
        \boldsymbol{E}_s (\boldsymbol{r}) \\
        \boldsymbol{H}_s (\boldsymbol{r})
    \end{pmatrix}
    + \sum_{n=1}^{n=N}
    \begin{pmatrix}
        \xi^2 \boldsymbol{G} (\boldsymbol{r}, \boldsymbol{r}_n) \boldsymbol{\alpha}_E & i \xi \boldsymbol{G}_{EH} (\boldsymbol{r}, \boldsymbol{r}_n) \boldsymbol{\alpha}_H \\
        -i \xi \boldsymbol{G}_{EH} (\boldsymbol{r}, \boldsymbol{r}_n) \boldsymbol{\alpha}_E & \xi^2 \boldsymbol{G} (\boldsymbol{r}, \boldsymbol{r}_n) \boldsymbol{\alpha}_H
    \end{pmatrix}
    \begin{pmatrix}
        \boldsymbol{E}(\boldsymbol{r}_n) \\
        \boldsymbol{H}(\boldsymbol{r}_n)
    \end{pmatrix}
    \label{eq:fields}
\end{equation}
where we have chosen units such that the impedance of free space is $\eta_0 = 1$ and work in terms of a dimensionless wavenumber $\xi$.
This solution is not yet closed, since the fields applied to each scatterer $(\boldsymbol{E} (\boldsymbol{r}_n), \boldsymbol{H} (\boldsymbol{r}_n))$ must be determined.
Imposing self--consistency yields the following matrix equations connecting the source and total fields at each scatterer
\begin{equation}
    \boldsymbol{R}_{nm} 
    \begin{pmatrix}
        \boldsymbol{E}(\boldsymbol{r}_m) \\
        \boldsymbol{H}(\boldsymbol{r}_m)
    \end{pmatrix}
    =
    \begin{pmatrix}
        \boldsymbol{E}_s (\boldsymbol{r}_n) \\
        \boldsymbol{H}_s (\boldsymbol{r}_n)
    \end{pmatrix} ,
    \label{eq:self-const}
\end{equation}
where 
\begin{equation}
    \boldsymbol{R}_{nm} = 
    \begin{pmatrix}
        \boldsymbol{1}\delta_{nm} - \xi^2 \boldsymbol{G} (\boldsymbol{r}_n, \boldsymbol{r}_m) \boldsymbol{\alpha}_E & - i \xi \boldsymbol{G}_{EH} (\boldsymbol{r}_n, \boldsymbol{r}_m) \boldsymbol{\alpha}_H \\
        i \xi \boldsymbol{G}_{EH} (\boldsymbol{r}_n, \boldsymbol{r}_m) \boldsymbol{\alpha}_E & \boldsymbol{1}\delta_{nm} - \xi^2 \boldsymbol{G} (\boldsymbol{r}_n, \boldsymbol{r}_m) \boldsymbol{\alpha}_H
    \end{pmatrix} .
\end{equation}
The self--consistency condition (\ref{eq:self-const}) can be solved with standard matrix methods \cite{NumericalRecipes} for the fields applied to each scatterer, which includes the source field as well as contributions from all of the other scatterers.
Once these fields are found, the solution to Maxwell's equations given by (\ref{eq:fields}) is fully specified.

The particular physical system we consider in numerical examples is an arrangement of silicon spheres of radius 65 nm at a wavelength of 550 nm giving $k r \approx 0.75$.
For this simple choice of metasurface element, the electric and magnetic polarisability tensors can be constructed from the Mie coefficients \cite{Mie1908, Bohren1983} $a_1$ and $b_1$ as
\begin{align}
    \boldsymbol{\alpha}_E &= \boldsymbol{1} i \frac{6 \pi}{k^3} a_1 & \boldsymbol{\alpha}_H &= \boldsymbol{1} i \frac{6 \pi}{k^3} b_1
\end{align}
where $\boldsymbol{1} = {\rm diag} (1,1,1)$ is the unit tensor.
Polarisability tensors for more complicated scatterers can be extracted from numerical modelling \cite{Arango2013, Liu2016}, making this method applicable to a very wide range of systems.
The key benefit is that an expensive full--wave simulation is required for only a single scatterer, not the entire metasurface.

\section{DESIGNING METASURFACES}
\label{sec:designing}

In the previous section, we derived expressions that give the effect of a metasurface, defined by a collections of scatterers at locations $\{ \boldsymbol{r}_n \}$ and with polarisabiltities $\boldsymbol{\alpha}_E$ and $\boldsymbol{\alpha}_H$.
The aim is to now find a way to choose the distribution of the scatterers $\{ \boldsymbol{r}_n \}$ to achieve a desired wave--scattering effect.
We consider how moving one of the scatterers by a small amount changes the fields.
Taylor expanding the Dirac--deltas in (\ref{eq:PM}) as
\begin{align}
    \delta (\boldsymbol{r} - \boldsymbol{r}_n - \delta \boldsymbol{r}_n) = \delta (\boldsymbol{r} - \boldsymbol{r}_n) + (\delta \boldsymbol{r}_n \cdot \nabla) \delta (\boldsymbol{r} - \boldsymbol{r}_n) + \frac{1}{2} ( \delta \boldsymbol{r}_n \cdot \nabla )^2 \delta (\boldsymbol{r} - \boldsymbol{r}_n) + \ldots
\end{align}
and keeping only the first order terms gives the variation in the field due to a small change in the position of the scatterers 
\begin{equation}
    \begin{pmatrix}
        \delta \boldsymbol{E} (\boldsymbol{r}) \\
        \delta \boldsymbol{H} (\boldsymbol{r})
    \end{pmatrix}
    = -
    \begin{pmatrix}
        \xi^2 \boldsymbol{G}(\boldsymbol{r}, \boldsymbol{r}_n) \boldsymbol{\alpha}_E & i \xi \boldsymbol{G}_{EH}(\boldsymbol{r}, \boldsymbol{r}_n) \boldsymbol{\alpha}_H \\
        - i \xi \boldsymbol{G}_{EH}(\boldsymbol{r}, \boldsymbol{r}_n) \boldsymbol{\alpha}_E & \xi^2 \boldsymbol{G}(\boldsymbol{r}, \boldsymbol{r}_n) \boldsymbol{\alpha}_H
    \end{pmatrix}
    \begin{pmatrix}
        \nabla \boldsymbol{E} (\boldsymbol{r}_n) \\
        \nabla \boldsymbol{H} (\boldsymbol{r}_n)
    \end{pmatrix} \delta \boldsymbol{r}_n ,
    \label{eq:field_var}
\end{equation}
where the fields are given by (\ref{eq:fields}).
This gives an expression for how moving the position of a single scatterer changes the fields at every point in space.
These expressions can be used to find how changing the location of a scatterer affects a given figure of merit, which is a functional of the field configurations $\mathcal{F}[\boldsymbol{E}, \boldsymbol{H}]$.
Moving one scatterer produces a small change in the fields at every point in space, which in turn changes the figure of merit by a small amount 
\begin{equation}
    \mathcal{F}[\boldsymbol{E}, \boldsymbol{H}] \rightarrow \mathcal{F}[\boldsymbol{E}, \boldsymbol{H}] + \delta \mathcal{F}[\boldsymbol{E}, \boldsymbol{H}, \delta \boldsymbol{E}, \delta \boldsymbol{H}] .
\end{equation}
The change in the figure of merit is linear in $\delta \boldsymbol{r}_n$, so once we have derived the analytic expression for $\delta \mathcal{F}$ it can be used to find an expression for a $\delta \boldsymbol{r}_n$ that leads to an increase in the figure of merit.
In this way, we can begin from an initial distribution of scatterers and iteratively calculate the set of moves for each scatterer $\{\delta \boldsymbol{r}_n\}$ that increase the figure of merit.
In the following examples we demonstrate the versatility of this procedure by considering three different figures of merit.

\begin{figure}[ht]
    \centering
    \includegraphics[width=\linewidth]{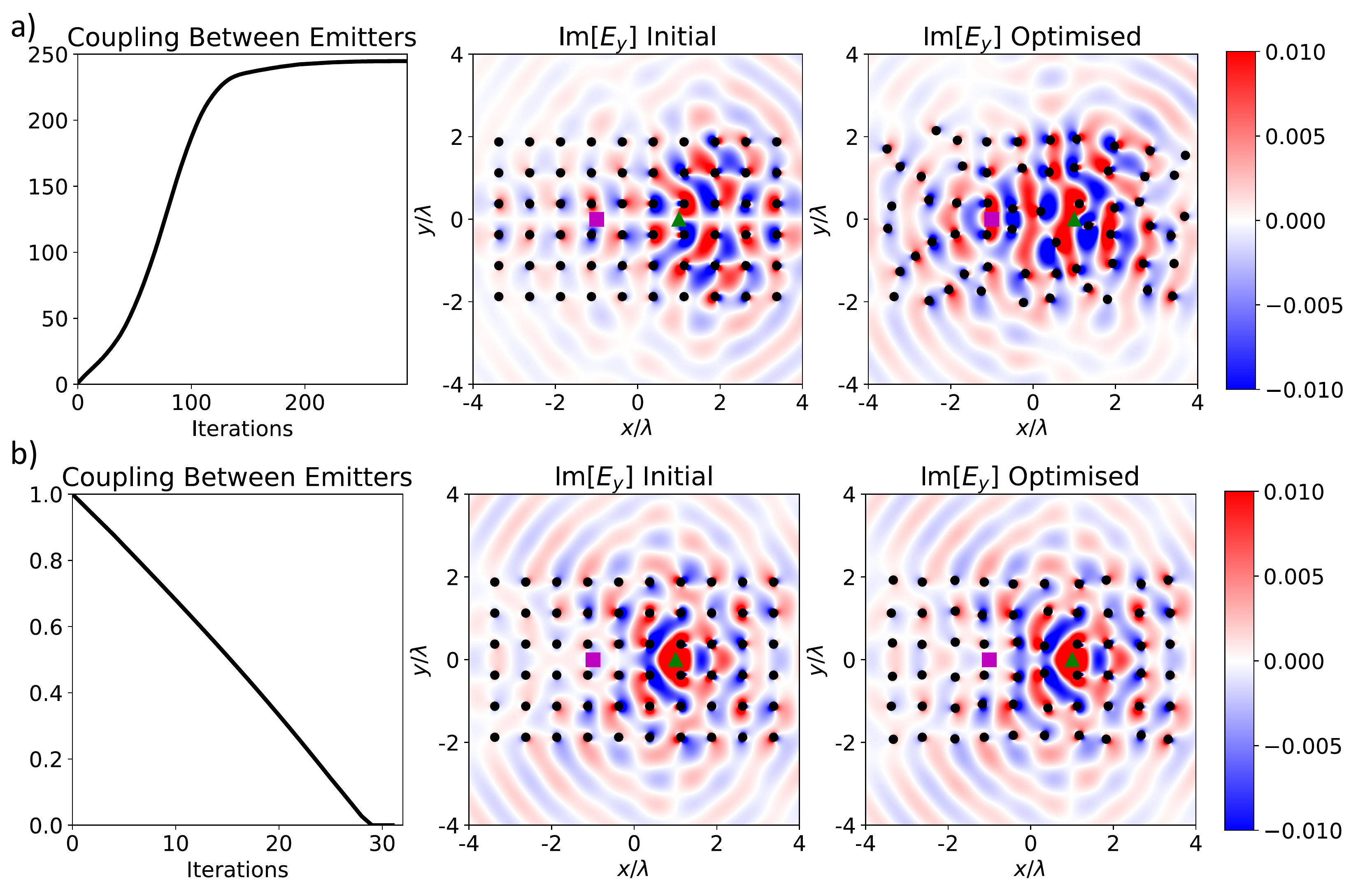}
    \caption{Designing dielectric structures that manipulate the coupling between two nearby emitters.  
    Beginning from an initial distribution of scatterers, shown in the centre panels, the update equation (\ref{eq:coupling_update}) is used to iteratively move the scatterers to a) increase and b) decrease the coupling between two nearby emitters.
    The change in coupling over the optimisation procedure is shown in the left--hand panels and the optimised structure is shown in the right--hand panels.
    The two emitters are shown as a magenta square and a green triangle.
    The polarisation of the emitters in a) is $\boldsymbol{p} = \hat{\boldsymbol{x}}$ for the green triangle and $\boldsymbol{p} = \boldsymbol{\hat{y}}$ for the magenta square, while in b) is $\boldsymbol{p} = \boldsymbol{\hat{y}}$ for both emitters.
    In both cases, the $y$ component of the field is re--shaped by moving the scatterers to exhibit either a null or peak at the location of the scatterer shown by the magneta square.}
    \label{fig:coupling}
\end{figure}
Firstly, we consider the coupling between two emitters with different polarisations.
For this problem, we have two sources located at $\boldsymbol{r}_{s,1}$ and $\boldsymbol{r}_{s,2}$ and with electric polarisations $\boldsymbol{p}_{1,2}$.
This means that the source fields in Maxwell's equations (\ref{eq:maxwell}) can be written as  
\begin{align}
    \boldsymbol{E}_s (\boldsymbol{r}) &= \xi^2 \boldsymbol{G} (\boldsymbol{r}, \boldsymbol{r}_{s,1}) \cdot \boldsymbol{p}_1 + \xi^2 \boldsymbol{G} (\boldsymbol{r}, \boldsymbol{r}_{s,2}) \cdot \boldsymbol{p}_2, \\
    \boldsymbol{H}_s (\boldsymbol{r}) &= - i \xi \boldsymbol{G}_{EH} (\boldsymbol{r}, \boldsymbol{r}_{s,1}) \cdot \boldsymbol{p}_1 - i \xi \boldsymbol{G}_{EH} (\boldsymbol{r}, \boldsymbol{r}_{s,2}) \cdot \boldsymbol{p}_2 ,
\end{align}
assuming that the sources are small compared to the wavelength.
The coupling between the two sources is then
\begin{equation}
    \rho_{12} = {\rm Im} \left[ \boldsymbol{p}^*_{1} \cdot \boldsymbol{E}_2 (\boldsymbol{r}_1) \right] .
\end{equation}
where $\boldsymbol{E}_2 (\boldsymbol{r}_1)$ is the field generated by the second emitter, along with the scattering structure, at the first emitter.
In this way, $\rho_{12}$ characterises the overlap of the fields generated by the emitters.
To design a structure that manipulates the coupling between two emitters, we expand the figure of merit to first order under small changes in the fields at the second emitter
\begin{align}
    \mathcal{F}_{\rm coupling} &= {\rm Im} \left[ \boldsymbol{p}^*_1 \cdot ( \boldsymbol{E}_2(\boldsymbol{r}_1) + \delta \boldsymbol{E}_2(\boldsymbol{r}_1)) \right] , \\
    \delta F_{\rm coupling} &= {\rm Im} \left[ \boldsymbol{p}^*_1 \cdot \delta \boldsymbol{E}_2(\boldsymbol{r}_1) \right] .
\end{align}
Inserting the expression for the variation of the fields (\ref{eq:field_var}), we find
\begin{equation}
    \delta F_{\rm coupling} = - \sum_n {\rm Im} \left[ \boldsymbol{p}^*_1 \cdot \left\{ \xi^2 \boldsymbol{G}(\boldsymbol{r}_1, \boldsymbol{r}_n) \boldsymbol{\alpha}_E \nabla \boldsymbol{E} (\boldsymbol{r}_n) + i \xi \boldsymbol{G}_{EH}(\boldsymbol{r}_1, \boldsymbol{r}_n) \boldsymbol{\alpha}_H \nabla \boldsymbol{H}(\boldsymbol{r}_n) \right\}  \right] \delta \boldsymbol{r}_n .
\end{equation}
This gives a way of calculating a move of the $n^{\rm th}$ scatterer so that the figure of merit is guaranteed to either increase or decrease.
Choosing 
\begin{equation}
    \delta \boldsymbol{r}_n \propto \mp {\rm Im} \left[ \boldsymbol{p}^*_1 \cdot \left\{ \xi^2 \boldsymbol{G}(\boldsymbol{r}_1, \boldsymbol{r}_n) \boldsymbol{\alpha}_E \nabla \boldsymbol{E} (\boldsymbol{r}_n) + i \xi \boldsymbol{G}_{EH}(\boldsymbol{r}_1, \boldsymbol{r}_n) \boldsymbol{\alpha}_H \nabla \boldsymbol{H}(\boldsymbol{r}_n) \right\}  \right]
    \label{eq:coupling_update}
\end{equation}
leads to a positive $\delta \mathcal{F}_{\rm coupling}$ if the negative sign is taken, and a negative $\delta \mathcal{F}_{\rm coupling}$ if the positive sign is taken.
This gives a way of calculating how to move all of the scatterers at the same time in a way that changes the figure of merit in the desired way.
An example of applying this procedure to change the coupling between two emitters is shown in Figure \ref{fig:coupling}.
In Figure \ref{fig:coupling}a, the coupling between an electric dipole with polarisation $\boldsymbol{p} = \boldsymbol{\hat{x}}$, shown as a green triangle, and an electric dipole with polarisation $\boldsymbol{p} = \boldsymbol{\hat{y}}$, shown as a magenta square, is enhanced.
The scatterer positions are updated according to the upper sign of (\ref{eq:coupling_update}), leading to a redistribution of the scattered field.
To increase the coupling, the $\boldsymbol{\hat{y}}$ component of electric field at the location of the emitter with polarisation $\boldsymbol{\hat{y}}$ is increased greatly.
Another case of interest might be to reduce the coupling between two similarly polarised nearby emitters.
Taking the lower sign in the update equation (\ref{eq:coupling_update}) and decreasing the coupling between two emitters with the same polarisation, $\boldsymbol{\hat{y}}$, is demonstrated in Figure \ref{fig:coupling}b.
The scatterers are now redistributed to place a null in the field at the location of the emitter shown by the magenta square.

\begin{figure}
    \centering
    \includegraphics[width=\linewidth]{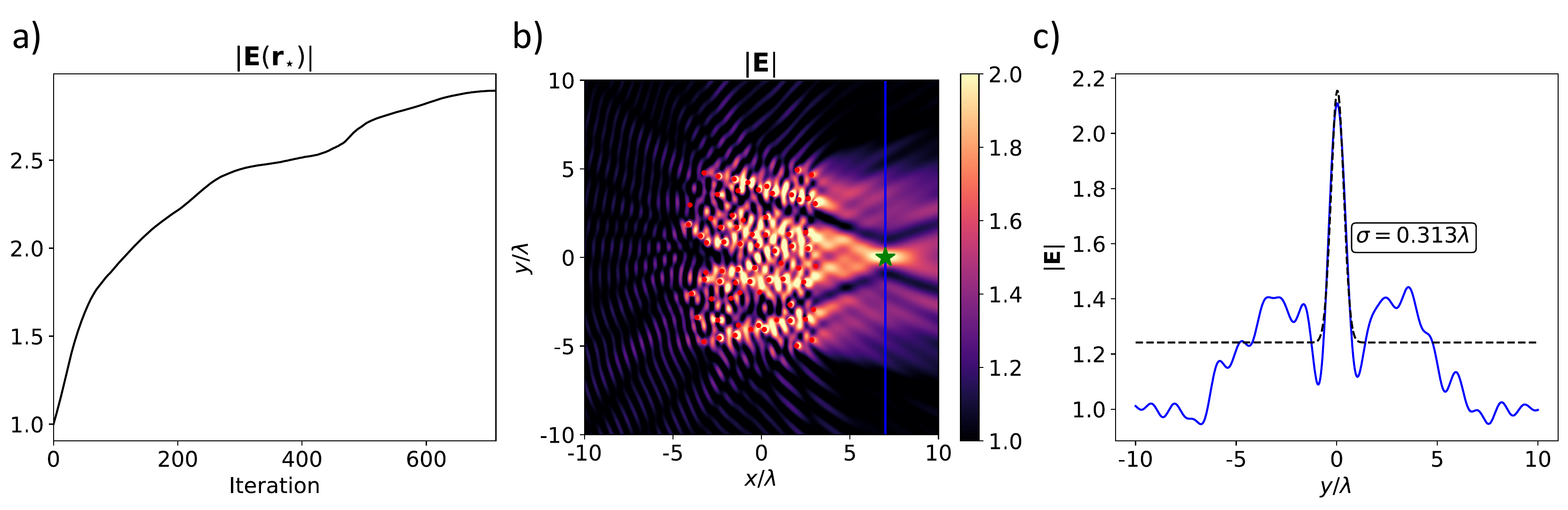}
    \caption{The design of a metamaterial than focuses the energy from a plane wave to a point, shown as the green star.  
    The figure of merit for this optimisation is the modulus of the electric field at the target location (\ref{eq:fom_lens}); a) shows the increase of this quantity as a function of progressing optimisation and b) is the final design. A cut of the field along the blue line is given in c) showing the narrow focus.}
    \label{fig:lensing}
\end{figure}
Secondly, we consider the problem of focusing a plane--wave to a point.
For this problem, the source fields in the solutions to Maxwell's equations (\ref{eq:fields}) are plane waves, with a particular polarisation and wave--vector.
For the example in Figure \ref{fig:lensing}, we choose a TE polarised wave with wave--vector $\boldsymbol{k} = k (1,0,0)$.
The figure of merit is the magnitude of the electric field at the target location $\boldsymbol{r}_\star$.
\begin{equation}
    \mathcal{F}_{\rm lens} = |\boldsymbol{E}(\boldsymbol{r}_\star)| .
    \label{eq:fom_lens}
\end{equation}
This can be expanded to first order under small changes in the fields as 
\begin{align}
    |\boldsymbol{E}(\boldsymbol{r}_\star)| &= \sqrt{\boldsymbol{E}(\boldsymbol{r}_\star) \cdot \boldsymbol{E}^*(\boldsymbol{r}_\star)} , \\
    &= \sqrt{(\boldsymbol{E}(\boldsymbol{r}_\star) + \delta \boldsymbol{E}(\boldsymbol{r}_\star))\cdot (\boldsymbol{E}^*(\boldsymbol{r}_\star) + \delta \boldsymbol{E}^*(\boldsymbol{r}_\star) )} , \\
    &= \sqrt{|\boldsymbol{E}(\boldsymbol{r}_\star)|^2 + 2 {\rm Re} \left[ \delta \boldsymbol{E} (\boldsymbol{r}_\star) \cdot \boldsymbol{E}^* (\boldsymbol{r}_\star) \right]} , \\
    &= |\boldsymbol{E}(\boldsymbol{r}_\star)| \sqrt{1 + \frac{2 {\rm Re} \left[ \delta \boldsymbol{E} (\boldsymbol{r}_\star) \cdot \boldsymbol{E}^* (\boldsymbol{r}_\star)\right]}{|\boldsymbol{E}(\boldsymbol{r}_\star)|^2}} , \\
    &\approx |\boldsymbol{E}(\boldsymbol{r}_\star)| + \frac{{\rm Re} \left[ \delta \boldsymbol{E}(\boldsymbol{r}_\star) \cdot \boldsymbol{E}^*(\boldsymbol{r}_\star)\right]}{|\boldsymbol{E}(\boldsymbol{r}_\star)|} \label{eq:mod_expansion}.
\end{align}
Substituting the expression for the variation of the fields gives the following change in the figure of merit $\mathcal{F}_{\rm lens}$
\begin{equation}
    \delta \mathcal{F}_{\rm lens} = \frac{-1}{|\boldsymbol{E} (\boldsymbol{r}_\star)|} \sum_n {\rm Re} \left[ \left\{ \xi^2 \boldsymbol{G}(\boldsymbol{r}_\star, \boldsymbol{r}_n) \boldsymbol{\alpha}_E \nabla \boldsymbol{E} (\boldsymbol{r}_n) + i \xi \boldsymbol{G}_{EH}(\boldsymbol{r}_\star, \boldsymbol{r}_n) \boldsymbol{\alpha}_H \nabla \boldsymbol{H}(\boldsymbol{r}_n) \right\} \cdot \boldsymbol{E}^* (\boldsymbol{r}_\star) \right] \delta \boldsymbol{r}_n .
\end{equation}
As this is linear in $\delta \boldsymbol{r}_n$, this gives a way of choosing $\delta \boldsymbol{r}_n$ so that the figure of merit increases.
The result of applying this procedure is shown in Figure \ref{fig:lensing}.
A structure is designed that focuses the field to the desired location.
Fitting a Gaussian of the form
\begin{equation}
    \mathcal{G}(y) = A \exp \left( \frac{(y-\mu)^2}{2 \sigma^2}\right) + B
\end{equation}
to the peak, we find that the width is $\sim \lambda /3$.

\begin{figure}
    \centering
    \includegraphics[width=\linewidth]{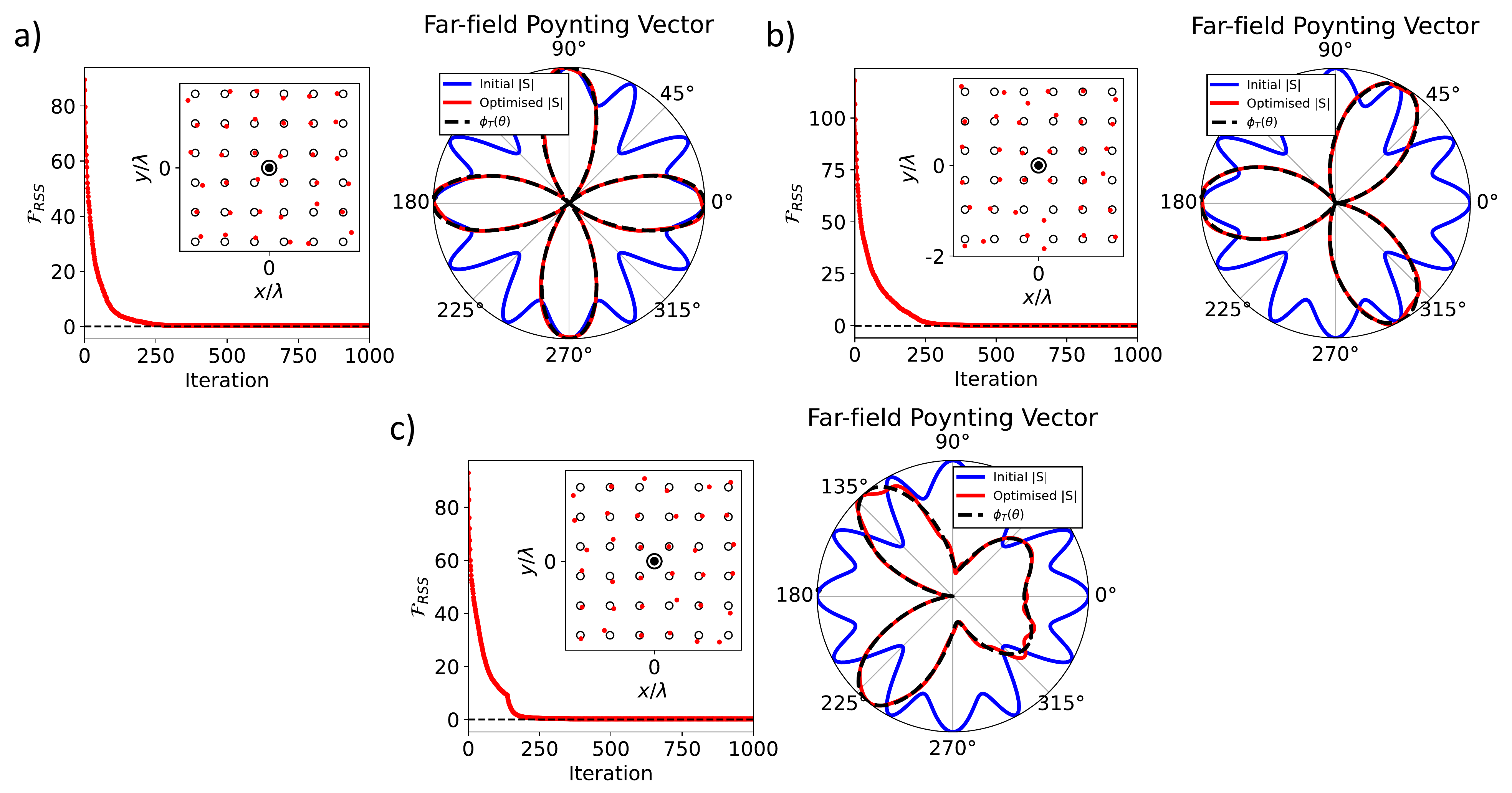}
    \caption{The design of a metamaterial with a chosen radiation pattern.
    In each case, the structure is driven by an emitter polarised along the $z$ axis at the origin.  
    For each of the target radiation patterns, black dashed lines, the scatterers begin at the locations indicated by black circles and are iteratively moved to reduce the difference between the radiation pattern and the desired pattern (\ref{eq:fom_rss}).
    The optimised locations of the scatterers are shown as red dots and the final radiation patterns as red lines.
    }
    \label{fig:farfield}
\end{figure}
The final problem we consider, is shaping the far--field Poynting vector of an emitter.
We would like $|\boldsymbol{S}(\theta)|$ to have a particular shape, $\phi_T (\theta)$ in the far--field.
Our aim is to design a scattering structure that produces a particular $|\boldsymbol{S}(\theta)|$ in the far--field, defined by a target angular distribution $\phi_T (\theta)$.
One way this can be achieved is by minimising the residual sum of squares between the current angular distribution of the Poynting vector and the target distribution
\begin{equation}
    \mathcal{F}_{\rm RSS} = \sum_i \left[ |\boldsymbol{S}(\theta_i)| - \phi_T(\theta_i) \right]^2 .
    \label{eq:fom_rss}
\end{equation}
In order to use this figure of merit, both $|\boldsymbol{S}(\theta_i)|$ and $\phi_T(\theta_i)$ must be normalised to range from 0 to 1.
It should be noted that the choice of figure of merit is not unique: one could seek to maximise the overlap integral between the current angular distribution of Poynting vector and the target distribution \cite{Capers2021}.
To derive an expression that can be used to calculate how the scatterers should be moved to minimise this figure of merit, we expand first under small changes in $\delta |\boldsymbol{S}(\theta)|$
\begin{align}
    \sum_i \left[ |\boldsymbol{S}(\theta_i)| - \phi_T(\theta_i) \right]^2 &= \sum_i \left( |\boldsymbol{S}(\theta_i)| + \delta |\boldsymbol{S}(\theta_i)| - \phi_T(\theta_i) \right) \left( |\boldsymbol{S}(\theta_i)| + \delta |\boldsymbol{S}(\theta_i)| - \phi_T(\theta_i) \right) , \\
    &= \sum_i |\boldsymbol{S}(\theta_i)|^2 + \phi_T^2 (\theta_i) + 2 \delta |\boldsymbol{S}(\theta_i)| (|\boldsymbol{S}(\theta_i)| - \phi_T (\theta_i)) ,
\end{align}
and retaining only first order terms, we find that 
\begin{equation}
    \delta \mathcal{F}_{\rm RSS} = \sum_i \left[ 2 \delta |\boldsymbol{S}(\theta_i)| (|\boldsymbol{S}(\theta_i)| - \phi_T (\theta_i) \right] .
    \label{eq:deltaFrss}
\end{equation}
It is then necessary to find $\delta |\boldsymbol{S}|$, the variation in the Poynting vector, in terms of the variations in the fields (\ref{eq:field_var}).
Using the expression we obtained from expanding the modulus of the electric field (\ref{eq:mod_expansion}), we know that
\begin{equation}
    \delta |\boldsymbol{S}| = \frac{{\rm Re} \left [\delta \boldsymbol{S} \cdot \boldsymbol{S}^* \right]}{|\boldsymbol{S}|} .
\end{equation}
Then, $\delta \boldsymbol{S}$ can be derived from the expression for the Poynting vector
\begin{align}
    \boldsymbol{S} &= \frac{1}{2} \boldsymbol{E} \times \boldsymbol{H}^* \\
    &= \frac{1}{2} (\boldsymbol{E} + \delta \boldsymbol{E}) \times (\boldsymbol{H}^* + \delta \boldsymbol{H}^*), \\
    \delta \boldsymbol{S} &= \frac{1}{2} \left[ \boldsymbol{E} \times \delta \boldsymbol{H}^* + \delta \boldsymbol{E} \times \boldsymbol{H}^* \right] .
\end{align}
Substituting this into (\ref{eq:deltaFrss}) gives us the change of the figure of merit in terms of the changes in the fields, which are linear in $\delta \boldsymbol{r}_n$.
As with the previous examples, the expressions for the field variations (\ref{eq:field_var}) can be substituted in to yield an expression for moving the scatterers to decrease this figure of merit.
Figure \ref{fig:farfield} shows several examples of using this process to design scatterering structures with arbitrary far--field radiation patterns.

\section{SUMMARY \& CONCLUSIONS}

We have derived a method of designing metamaterials built from several discrete scatterers that exhibit electric and magnetic dipole resonances.
While gradient based, our method leverages the advantages of the adjoint method and by utilising the discrete dipole approximation to avoid full--wave simulations ensures numerical efficiency.
We have applied our design methodology to three different problems, relevant to nanophotonics.
The coupling between two nearby emitters can be manipulated with an appropriate photonic structure to increase coupling by a factor of ~250 or massively reduce the coupling, removing cross--talk.
A plane wave can be focused to a chosen point, with a focus width of $\sim \lambda / 3$.
Finally, we demonstrate the design of a dielectric antenna with any desired radiation pattern.

This framework might be extended beyond the dipole approximation to include high order multipoles, to achieve more diverse control of light.
Developing the method to design structures that perform differernt functions for different exciting fields would be very useful for optical computing applications.
The general idea of analytically expanding figures of merit under small perturbations in design parameters to find efficient ways to calculate gradients could be applied to many other optics problems, from fibre optics to imaging through disorder.

\acknowledgments 
 
We acknowledge financial support from the Engineering and Physical Sciences Research Council (EPSRC) of the United Kingdom, via the EPSRC Centre for Doctoral Training in Metamaterials (Grant No. EP/L015331/1). 
J.R.C also wishes to acknowledge financial support from Defence Science Technology Laboratory (DSTL). 
S.A.R.H acknowledges financial support from the Royal Society (RPG-2016-186).

\copyright Copyright 2022 Society of Photo‑Optical Instrumentation Engineers (SPIE).
According to SPIE Article-Sharing Policies ``Authors may post draft manuscripts on preprint servers such as arXiv. 
If the full citation and Digital Object Identifier (DOI) are known, authors are encouraged to add this information to the preprint record.'' \url{https://www.spiedigitallibrary.org/article-sharing-policies}.

This document represents a draft from J. R. Capers, S. J. Boyes, A. P. Hibbins and S. A. R. Horsley ``Designing Metasurfaces to Manipulate Antenna Radiation'', Proc. SPIE 12130, Metamaterials XIII, 121300H (24 May 2022); \url{https://doi.org/10.1117/12.2621160}
Please check out the SPIE paper for a complete list of figures, tables, references and general content.


\end{document}